# Optimizing the concentration ratio of multi-faceted focusing heliostats

François Hénault[a], Gilles Flamant[b], Cyril Caliot[c]

(a) Institut de Planétologie et d'Astrophysique de Grenoble, Université Grenoble-Alpes, CNRS, B.P. 53, 38041 Grenoble – France
(b) Processes, Materials and Solar Energy laboratory, PROMES CNRS, 7 Rue du Four Solaire, 66120 Font-Romeu-Odeillo-Via – France
(c) CNRS, UPPA, E2S, LMAP, 1 Allée du parc Montaury, 64600 Anglet, France

## ABSTRACT

This technical note aims at optimizing the concentration ratio of multi-faceted focusing heliostats implemented into a solar tower power plant. The ideal shape of a heliostat located off-axis in the field is known to be the local section of a fictitious paraboloïd whose parameters are varying continuously with the Sun angular position. We describe an optimization procedure applicable to those heliostats. The flux densities formed at the solar receiver and the achievable concentrating ratios are computed using an improved convolution algorithm. It is shown that the optimized heliostat shape can produce typical concentration gains of approximately 10%, even when the heliostats reflect the Sun under large incidence angles.

## 1    INTRODUCTION

It is well known that the ideal shape of a focusing heliostat in a solar tower power plant is the local section of a fictitious paraboloïd whose focus is located at the centre of the solar receiver, and the optical axis is parallel to the Sun vector **S** at a given time [1]. Consequently, the ideal shape of the heliostat changes continuously with the time of the day and the day of the year. This drawback may be removed by defining a "Sun reference position" $S_0$ from which the heliostat parameters are fixed. Such improvement only involves slight re-alignments of the tilt angles of the heliostat mirrors around the horizontal and vertical axes, so that they become tangent to the ideal paraboloïd shape. Here is described an optimization procedure applicable to multi-faceted focusing heliostats (Section 2). The flux densities formed at the solar receiver and the achievable concentrating ratios are computed using an improved convolution algorithm (Section 3). It is shown that the optimized heliostat shape can produce gains of approximately 10% in terms of concentration ratio. A brief conclusion is drawn in Section 4.

## 2 PRINCIPLE

### 2.1 Solar tower plant configuration

Let us consider the case of a solar tower power plant whose general configuration is depicted in Figure 1-A. Two main coordinate systems are defined:

- The X'Y'Z' reference frame attached to the solar receiver with X'-axis directed from South to North, Y'-axis from East to West, and Z'-axis from Nadir to Zenith,
- The XYZ reference frame attached to an individual heliostat with X its optical axis and YZ its lateral dimensions along which its geometry is defined (see Figure 1-B and Table 1).
- Three vectors are defined in the X'Y'Z' reference frame (Figure 1-A) **S** is a unitary vector directed to the centre of the moving Sun,
- **R** is the unitary target vector directed from the heliostat centre to the solar receiver,
- **N** is the bisecting vector between both previous ones.

The vectors **S**, **R** and **N** obey the Snell-Descartes law for reflection that writes in vectorial form as:

$$\mathbf{S} + \mathbf{R} = 2(\mathbf{S}\ \mathbf{N})\mathbf{N} = 2\cos i\ \mathbf{N}, \qquad (1)$$

with $i$ the Sun incidence angle. The main employed parameters are summarized in Table 1. We consider the case of a heliostat located at coordinates (86.6, 50., 0.) expressed in meters into the X'Y'Z' reference frame. It may be noted that the distance $d$ from the heliostat to the solar receiver is kept equal to 100 meters and that the heliostat and the solar receiver are located at the same altitude along the Z'-axis, which is considered as the worst and most demanding case. The heliostat is made of $m \times n$ identical spherical modules of focal length $f = d = 100$ m. This is a simplified version of the focusing heliostats equipping the solar tower power plant in Targasonne, France.

Table 1: Main parameters of the solar power plant and of the focusing heliostat.

| Parameter | Symbol | Value | Unit |
|---|---|---|---|
| Target vector from heliostat to receiver | **R** | (86.6, 50., 0.) | m |
| Distance from heliostat to receiver | $d$ | 100 | m |
| Incidence angle on solar receiver | $\beta$ | 30 | degrees |
| Heliostat width along Y-axis | $w$ | 3.4 | m |
| Heliostat height along Z-axis | $h$ | 3. | m |
| Number of heliostat modules | $m \times n$ | 4 x 2 | |
| Module width along Y-axis | $w_M$ | 0.7 | m |
| Module height along Z-axis | $h_M$ | 1.4 | m |
| Module focal length | $f$ | $80 \leq f \leq 120$ | m |
| Solar receiver diameter | $d'$ | 1.2 | m |
| Mean Sun angles in azimuth and height | $(a_0, h_0)$ | (0., 44.63) | degrees |
| Mean Sun incidence angle | $i_0$ | 25.98 | degrees |

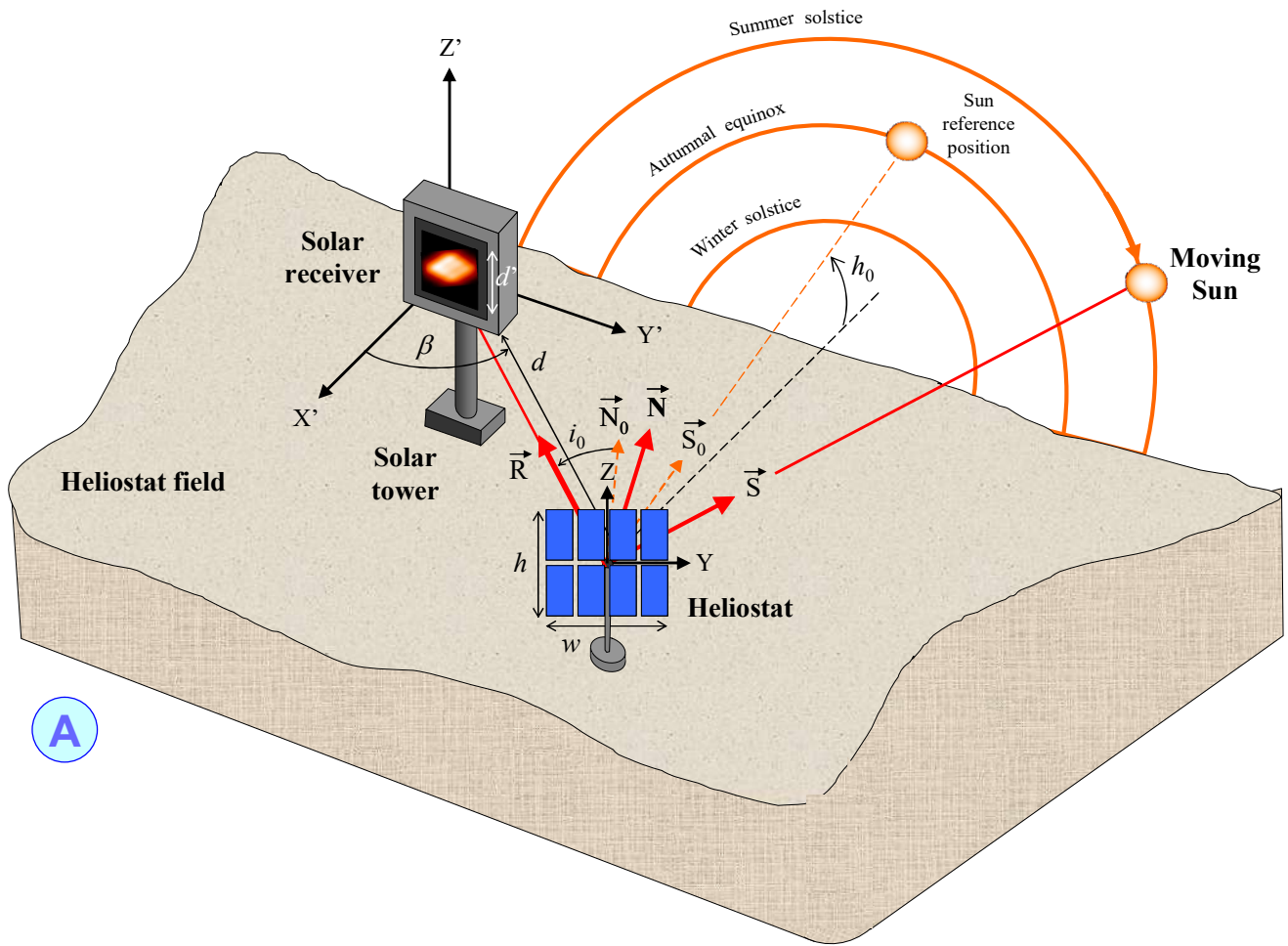

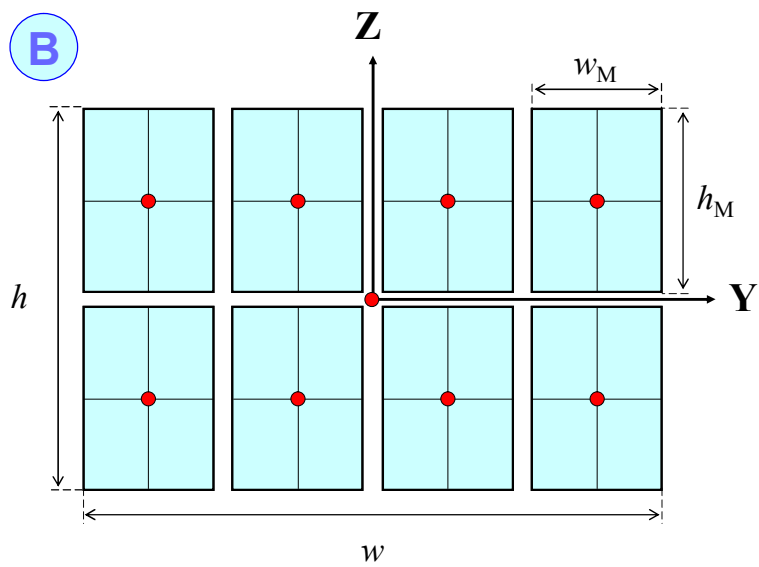

Figure 1: Solar tower power plant configuration (A). The geometry of the heliostats is shown on the bottom scheme (B).

## 2.2 Optimized off-axis heliostat

The optimized shape of the heliostat is named "off-axis" here after. It differs from the classical "spherical" shape, where the tilt angles of the modules around the Y and Y axes are adjusted in order to coincide with a monolithic sphere of focal length $f$ equal to the distance $d = 100$ m separating the heliostat from the solar receiver. Here it is assumed that all heliostat modules are identical. Then the sole degrees of freedom available for optimizing the off-axis heliostat are the tilt angles of each individual module around the Y and Z axes. The employed optimization procedure is as follows:

1. We firstly define a "Sun reference position" that is assumed to be an averaged position all over the year. It is assumed to be reached at noon on the autumnal equinox day. It corresponds to the Sun reference vector $\mathbf{S_0}$ plotted in Figure 1.

2. Knowing both $\mathbf{S_0}$ and the target vector $\mathbf{R}$ (that is unchanged) enables determining the unitary vector $\mathbf{N_0}$ normal to the heliostat for that Sun position, by inversion of Eq. 1 it comes

$$\mathbf{N_0} = (\mathbf{S_0} + \mathbf{R}) / \sqrt{2(1 + \mathbf{S_0}\mathbf{R})}. \tag{2}$$

Then the reference incidence angle on the heliostat is equal to $i_0 = \arccos(S_0 N_0)$.

3. From the knowledge of vectors $\mathbf{S_0}$, $\mathbf{N_0}$ and the incidence angle $i_0$; the tilt angles $a_{i,j}$ and $h_{i,j}$ of each heliostat module are evaluated using a set of analytical formulas defined by Eqs. 3. These formulas are strictly equivalent to those presented in Ref. [1]. Alternatively, these angles could be determined with the help of standard ray-tracing software such as Zemax™.

$$\begin{aligned} a_{i,j} &= \frac{y_{i,j}}{2d}\left(\frac{\cos^2 i_0 \cos^2 \phi + \sin^2 \phi}{\cos i_0}\right) - \frac{z_{i,j}}{4d}\frac{\sin 2\phi \sin^2 i_0}{\cos i_0} \\ h_{i,j} &= -\frac{y_{i,j}}{4d}\frac{\sin 2\phi \sin^2 i_0}{\cos i_0} + \frac{z_{i,j}}{2d}\left(\frac{\cos^2 i_0 \sin^2 \phi + \cos^2 \phi}{\cos i_0}\right) \end{aligned} \tag{3}$$

where $y_{i,j}$ and $z_{i,j}$ are the coordinates of each module centre and $\phi = \arctan(s_{0Y}, s_{0Z})$ with $(s_{0Y}, s_{0Z})$ the cosine directors of the reference Sun vector $\mathbf{S_0}$ along the Y and Z axes. All of them are expressed into the local heliostat reference frame XYZ.

4. Finally, the flux density maps formed by the off-axis heliostat in the receiver plane Y'Z' are computed with a double FFT algorithm described in Ref. [2].

## 3 NUMERICAL RESULTS

The values of the optimized angles $a_{i,j}$ and $h_{i,j}$ are given in Table 2 for each heliostat module, and compared with those of the spherical heliostat. The flux densities formed at the solar receiver are computed using an improved convolution algorithm for both the spherical and off-axis heliostat cases. They are illustrated by false-colour views in Figure 2. The angular radiance law of the Sun was assumed to follow Jose's formulas [3]. Cross-checking these results with those obtained using a Grid ray-tracing (GRT) model leads to RMS error differences about 1%, which are comparable to those presented in Ref. [2]. Here two different cases are distinguished:

A. Case of one single heliostat located at the coordinates (86.6, 50., 0.) expressed in meters into the X'Y'Z' reference frame,

B. Case of a couple of heliostats being symmetric with respect to the X'-axis and located respectively at the coordinates (86.6, 50., 0.) and (86.6, -50., 0.) meters into the X'Y'Z' reference frame. Then the flux density maps formed by each heliostat are simply added one to the other. The case B is the most commonly encountered since heliostat fields generally present a symmetry with respect to the X'-axis. The achieved concentrating ratios by the spherical and off-axis heliostats are presented in Table 3 for both cases A. and B. It shows a net advantage of about 10 % in terms of concentrating power for the off-axis heliostats. This gain occurs around half time of the day, typically from 10h00 to 14h00 GMT.

Table 2: Tilt angles of the spherical and off-axis heliostat modules and their relative differences.

| Indices i, j | Spherical heliostat | | Off-axis heliostat | | Angles difference | | Unit |
|---|---|---|---|---|---|---|---|
| | Tilt wrt Z $a_{i,j}$ | Tilt wrt Y $h_{i,j}$ | Tilt wrt Z $a_{i,j}$ | Tilt wrt Y $h_{i,j}$ | Tilt wrt Z $a_{i,j}$ | Tilt wrt Y $h_{i,j}$ | |
| 1, 1 | 12,75 | 7,50 | 14,19 | 8,34 | 1,44 | 0,84 | mrad |
| 2, 1 | 4,25 | 7,50 | 5,20 | 7,55 | 0,95 | 0,05 | mrad |
| 3, 1 | -4,25 | 7,50 | -3,85 | 6,78 | 0,40 | -0,72 | mrad |
| 4, 1 | -12,75 | 7,50 | -12,94 | 6,04 | -0,19 | -1,46 | mrad |
| 1, 2 | 12,75 | -7,50 | 12,75 | -5,89 | 0,00 | 1,61 | mrad |
| 2, 2 | 4,25 | -7,50 | 3,80 | -6,70 | -0,45 | 0,80 | mrad |
| 3, 2 | -4,25 | -7,50 | -5,19 | -7,49 | -0,94 | 0,01 | mrad |
| 4, 2 | -12,75 | -7,50 | -14,24 | -8,24 | -1,49 | -0,74 | mrad |

Table 3: Achieved concentration ratios by both the spherical and off-axis heliostats. Top rows: case of the single heliostat. Bottom rows: case of two heliostats being symmetric with respect to the X'-axis.

| Concentration ratio | 09-23-2022, Day time GMT | | | | |
|---|---|---|---|---|---|
| | T = 09h00 | T = 10h30 | T = 12h00 | T = 13h30 | T = 15h00 |
| Spherical heliostat x 1 | 38,1 | 37,7 | 32,9 | 16,4 | 6,5 |
| Spherical heliostat x 2 | 44,6 | 54,1 | 65,9 | 54,1 | 44,6 |
| Off-axis heliostat x 1 | 32,0 | 35,7 | 35,7 | 24,3 | 6,5 |
| Off-axis heliostat x 2 | 38,5 | 60,0 | 71,4 | 60,0 | 38,5 |

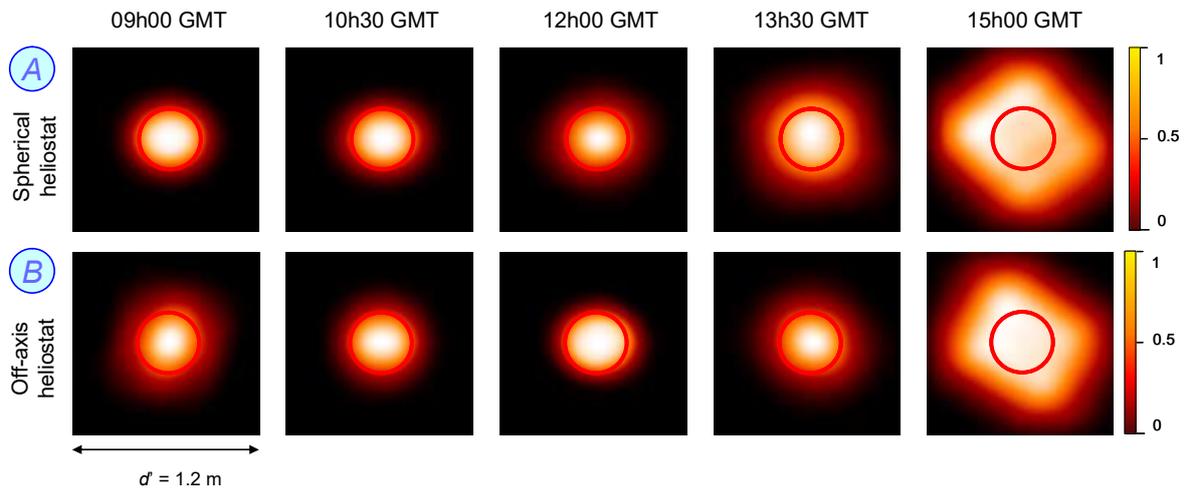

Figure 2: Flux densities formed at the solar receiver. (A) Case of spherical heliostat. (B) Case of the optimized off-axis heliostat. Red circles indicate the diameter of the ideally focused Sun image.

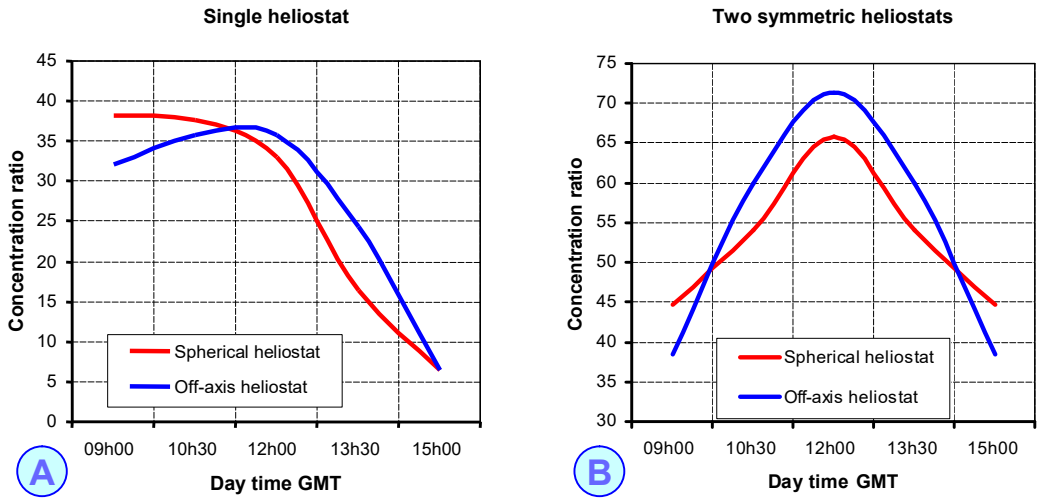

Figure 3: Achieved concentration ratios by both the spherical and off-axis heliostats. (A) Case of one single heliostat. (B) Case of two symmetric heliostats with respect to the X'-axis.

## 4  CONCLUSION

This short contribution considers the case of a multi-faceted heliostat focusing sunrays at the central receiver of a solar tower power plant. It presents a solution to improve the concentrating ratio of the heliostat in Sun-tracking mode all over daytime operation. The optimization process consists in turning the shape of a classical spherical heliostat into an off-axis shape profile. Assuming that all heliostat modules are identical, the available degrees of freedom for optimizing the spherical heliostat are the tilt angles of each of its individual modules. The optimization procedure firstly defines a Sun reference position on the sky, then slightly modifies these angles so that they are become tangent to an ideal parabolic section. A Fourier transform convolution model is used to evaluate the irradiance maps at the solar receiver and the achieved concentration ratios. Such an "off-axis" solution enables increasing the concentrating ratio of the heliostats by about 10 %. This procedure may be extended to the entire heliostat field, thus maximizing its concentration power at the solar receiver.